\documentclass[aps,prl,preprint,superscriptaddress]{revtex4-2}

\usepackage[english]{babel}
\usepackage{amsmath}
\usepackage{graphicx}
\usepackage{xcolor}
\usepackage[export]{adjustbox}
\usepackage{siunitx}
\usepackage{url}
\usepackage[colorlinks=true,allcolors=blue]{hyperref}
\begin{document}

\title{Terahertz frequency upconversion by coherently driving charge dynamics in the InSb/CdTe heterostructure}

\author{Pai Peng}
\thanks{These authors contributed equally to this work.}
\affiliation{State Key Laboratory of Low-Dimensional Quantum Physics and Department of Physics, Tsinghua University, Beijing 100084, China}
\affiliation{Beijing National Laboratory for Condensed Matter Physics, Institute of Physics, Chinese Academy of Sciences, Beijing 100190, China}

\author{Mingxiang Pan}
\thanks{These authors contributed equally to this work.}
\affiliation{School of Physics, Peking University, Beijing 100871, China}

\author{Jiuming Liu}
\thanks{These authors contributed equally to this work.}
\affiliation{School of Information Science and Technology, ShanghaiTech University, Shanghai 201210, China}

\author{Yi Yang}
\affiliation{Beijing National Laboratory for Condensed Matter Physics, Institute of Physics, Chinese Academy of Sciences, Beijing 100190, China}

\author{Lei Wang}
\affiliation{Beijing National Laboratory for Condensed Matter Physics, Institute of Physics, Chinese Academy of Sciences, Beijing 100190, China}

\author{Hao Lin}
\affiliation{State Key Laboratory of Low-Dimensional Quantum Physics and Department of Physics, Tsinghua University, Beijing 100084, China}

\author{Zehao Hu}
\affiliation{State Key Laboratory of Low-Dimensional Quantum Physics and Department of Physics, Tsinghua University, Beijing 100084, China}

\author{Jianlin Luo}
\affiliation{Beijing National Laboratory for Condensed Matter Physics, Institute of Physics, Chinese Academy of Sciences, Beijing 100190, China}
\affiliation{School of Physical Sciences, University of Chinese Academy of Sciences, Beijing 100049, China}

\author{Tao Dong}
\affiliation{International Center for Quantum Materials, School of Physics, Peking University, Beijing 100871, China}

\author{Xufeng Kou}
\email{kouxf@shanghaitech.edu.cn}
\affiliation{School of Information Science and Technology, ShanghaiTech University, Shanghai 201210, China}
\affiliation{ShanghaiTech Laboratory for Topological Physics, ShanghaiTech University, Shanghai 201210, China}

\author{Xinbo Wang}
\email{xinbowang@iphy.ac.cn}
\affiliation{Beijing National Laboratory for Condensed Matter Physics, Institute of Physics, Chinese Academy of Sciences, Beijing 100190, China}
\affiliation{School of Physical Sciences, University of Chinese Academy of Sciences, Beijing 100049, China}

\author{Huaqing Huang}
\email{huaqing.huang@pku.edu.cn}
\affiliation{School of Physics, Peking University, Beijing 100871, China}
\affiliation{Center for High Energy Physics, Peking University, Beijing 100871, China}
\affiliation{Collaborative Innovation Center of Quantum Matter, Beijing 100871, China}

\author{Luyi Yang}
\email{luyi-yang@mail.tsinghua.edu.cn}
\affiliation{State Key Laboratory of Low-Dimensional Quantum Physics and Department of Physics, Tsinghua University, Beijing 100084, China}
\affiliation{Frontier Science Center for Quantum Information, Beijing 100084, China}

\date{\today}

\begin{abstract}
We investigate terahertz (THz) harmonic generation in the InSb/CdTe heterostructure, demonstrating, for the first time, efficient in-plane magnetic field-induced second-harmonic generation (SHG). We also achieve significant third-harmonic generation (THG), rivalling Dirac materials such as graphene and Cd$_3$As$_2$. Our theoretical analysis identifies the primary SHG mechanism as the orbital-Zeeman correction to Drude conductivity, while the dominant THG contribution also shows Drude-like behavior. The results provide a general route to efficient THz harmonic generation in high mobility materials.
\end{abstract}

\maketitle

\section{Introduction}

The terahertz (THz) frequency range (0.1--several THz) is of central importance for next-generation wireless   communications, spectroscopy, and sensing, driving strong demand for compact and efficient on-chip THz sources.\cite{2018_Terahertz_NE,2022_Terahertz_ITC} A key challenge in this field, particularly for multiplier-chain-based complementary metal-oxide-semiconductor (CMOS) sources, is the urgent need for novel frequency upconversion schemes that offer high efficiency and wide bandwidth.\cite{2022_Terahertz_ITC} In addition to advancements in device engineering, recent breakthroughs in fundamental science have unveiled new mechanisms for frequency upconversion in the DC limit. Second- and third-order nonlinear transport in quantum materials has garnered great attention in recent years.\cite{2017_Bulk_NP,2019_Observation_N,2021_Thirdorder_NNb,2023_Quantum_S,2023_Quantummetricinduced_N,2021_Nonlinear_NRP} This is not only because it provides insights into unique features such as symmetry breaking and quantum geometry,\cite{2018_Nonreciprocal_NC,2015_Quantum_PRL,2024_Unification_PRL,2024_Nonreciprocal_ARCMP,2025_Quantum_S} but also because it holds potential for energy harvesting and THz receivers.\cite{2020_Highfrequency_SA,2021_Roomtemperature_NN} While considerable progress has been made in understanding both intrinsic mechanisms---such as Berry curvature dipoles\cite{2015_Quantum_PRL} and Berry connection polarizability\cite{2021_Thirdorder_NNb,2014_Field_PRL,2015_Geometrical_PRB}---and extrinsic scattering effects,\cite{2019_Disorderinduced_NC,2024_Odd_PRL,2024_Quantum_PRBb} the question of whether these nonlinear transport concepts can be effectively generalized to the THz range remains largely unexplored. Addressing this gap could pave the way for transformative advancements in THz technology.

Recently, THz harmonic generation has emerged as a rapidly growing field. High-efficiency THz third-harmonic generation (THG) has been observed in high-mobility Dirac semimetals such as graphene\cite{2018_Extremely_N} and Cd$_3$As$_2$\cite{2020_Nonperturbative_NC,2020_Efficient_PRL}. THG has also proven to be a valuable tool for exploring collective modes in superconductors,\cite{2014_Lightinduced_S} topological surface states,\cite{2016_Strong_NC,2021_Terahertz_nQM}
and quasiparticle dynamics in correlated oxides.\cite{2024_Strong_PRL,2024_Terahertz_a}
Although THG is not constrained by inversion symmetry, efficient THz THG in materials remains rare. On the other hand, THz second-harmonic generation (SHG) has been observed in superconductors through nonequilibrium supercurrents\cite{2020_Nonreciprocal_PRL,2020_Terahertz_PRLa}
 and spintronic THz devices.\cite{2025_Ultrafast_NC,2023_Efficient_NC}
 However, compared to THG, SHG exhibits lower efficiency and limited tunability. While recent work has shown THz SHG in arc-shaped bismuth stripes\cite{2024_Tunable_NEa} and in Bi$_2$Se$_3$,\cite{2024_Observation_PRB} its efficiency is limited by the small active light-matter interaction area. Thus, material systems capable of supporting both even and odd harmonic orders in the terahertz regime are not yet well-investigated. The relatively unexplored area of SHG requires new insights into its underlying mechanisms, such as the newly discovered ultrafast spintronic effects,\cite{2025_Ultrafast_NC,2023_Efficient_NC}
to drive further progress.

Here we present the observation of both THz SHG and THG from a high-quality InSb epitaxial layer grown on a CdTe buffer layer. We report a pronounced in-plane-magnetic-field-induced THz SHG that scales as ${\sim}E^2B$. Through semiclassical wave-packet and Boltzmann transport frameworks, we attribute the observed THz SHG to magnetic-field-induced Zeeman corrections to the Drude conductivity, facilitated by spin-orbit coupling (SOC). Remarkably, THG in InSb exhibits an exceptionally large nonlinear coefficient, comparable to benchmark Dirac materials such as graphene and Cd$_3$As$_2$. Within the same theoretical framework, the THG signal also originates from Drude-type intraband nonlinear charge dynamics and necessitates strong SOC. These light-driven nonlinear dynamics enable picosecond-scale control of charge transport for ultrafast optoelectronic applications and can be extended to other narrow-bandgap, noncentrosymmetric semiconductors with large SOC\cite{2002_Study_JoAP} and systems exhibiting quantum geometric nonlinear effects.\cite{2021_Thirdorder_NNb,2024_Unification_PRL} Practically, the InSb/CdTe heterostructure provides a wafer-scale, CMOS-compatible platform for on-chip THz applications.

\section{Sample and experimental setup}
We investigate the nonlinear terahertz response of a high-quality $n$-type InSb (111) epitaxial film grown on an insulating CdTe (111) buffer layer on a semi-insulating GaAs substrate,\cite{2020_Epitaxial_APL}
hereafter denoted as InSb/CdTe/GaAs. For comparison, control samples consisting of InSb grown directly on GaAs (InSb/GaAs) and CdTe grown on GaAs (CdTe/GaAs) were studied under identical conditions. The InSb/CdTe/GaAs heterostructure exhibits higher mobility than InSb/GaAs, while CdTe/GaAs is insulating [Supplemental Material (SM) Note 1].

Both InSb and CdTe possess a bulk point group symmetry of $4\bar{3}m$, while the point group symmetry at the interface is $3m$. InSb is a narrow-bandgap semiconductor with a bandgap of $\sim 0.2~\mathrm{eV}$, while CdTe is a wide-bandgap semiconductor with a bandgap of $\sim 1.5~\mathrm{eV}$. The InSb/CdTe interface forms a type-I heterojunction with significant band bending. This structure creates a strong built-in electric field that breaks inversion symmetry at the interface,\cite{2023_RoomTemperature_AM} thereby resulting in Rashba SOC. In addition to this structural inversion asymmetry-induced Rashba SOC, bulk inversion asymmetry inherent to InSb leads to Dresselhaus SOC.

Based on a similar InSb/CdTe/GaAs sample, our previous nonlinear transport measurements have demonstrated a significant nonreciprocal charge transport response:\cite{2023_RoomTemperature_AM} when an external current is perpendicular to the in-plane magnetic field, a second-order nonlinear transport signal emerges along the direction of the external current. These findings motivate further investigation into the nonlinear response at THz frequencies. In addition to the second-order response, the third-order nonlinear response of high-mobility electrons in the narrow-bandgap semiconductor InSb is also of significant interest.\cite{2024_Third_OLO,2024_Complex_OLO}

\begin{figure}
  \centering
  \includegraphics[max width=\linewidth]{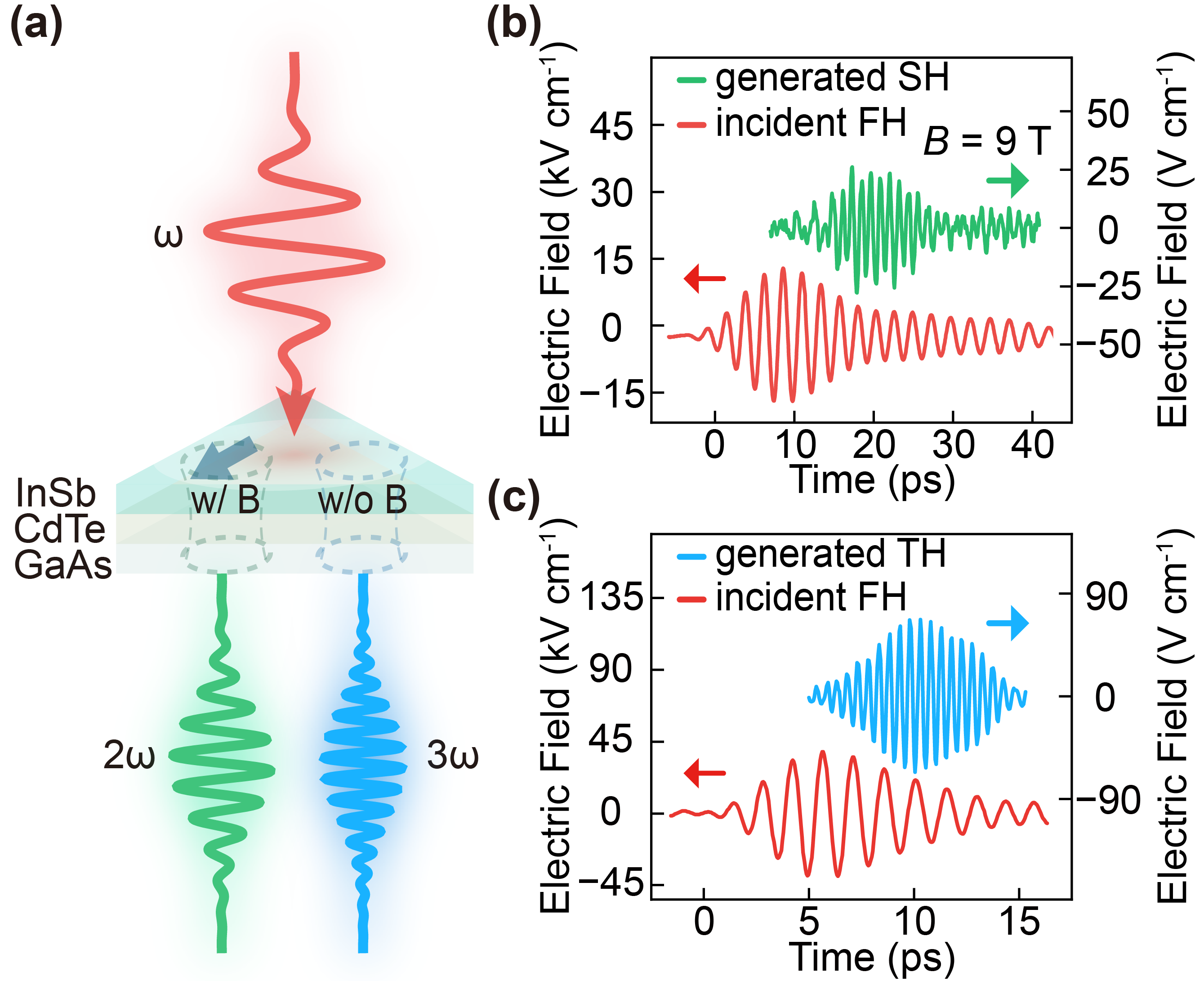}
  \caption{THz SHG and THG. (a) Schematic of THz frequency upconversion in InSb/CdTe/GaAs. A multicycle THz pulse ($\omega$) generates SHG under an in-plane magnetic field (b) and THG in the absence of a magnetic field (c).}
  \label{fig:boat1}
\end{figure}

Figure \ref{fig:boat1}(a) illustrates the schematic of the THz frequency upconversion processes in our sample. The multicycle fundamental THz radiation with central frequency $\omega$ is focused onto the sample. Under appropriate conditions, the InSb/CdTe heterostructure exhibits both second- and third-order responses [Fig. \ref{fig:boat1}(b,c)], denoted as 2$\omega$ and 3$\omega$, respectively, which will be discussed in detail. These measurements were performed using a home-built nonlinear THz spectroscopy setup capable of applying magnetic fields up to 10 T and cooling to 1.5 K.\cite{2024_Tabletop_RoSI} Most data were taken at 5 K, unless otherwise stated.

\section{Magnetic-field-induced SHG}
In our SHG experiments, the temporal waveform of the incident fundamental harmonic (FH) THz pulse centered at 0.42 THz is shown in Fig. \ref{fig:boat1}(b) (left axis). A band-pass filter centered at 0.84 THz was placed after the sample. First, we did not use polarizers in the detection path. As a result, the collected SHG signal contains all possible polarization components (SM Note 2). After mapping the time-domain traces using electro-optic sampling, we performed Fourier transforms to obtain the frequency-domain spectra. 

\begin{figure}
  \centering
  \includegraphics[max width=\linewidth]{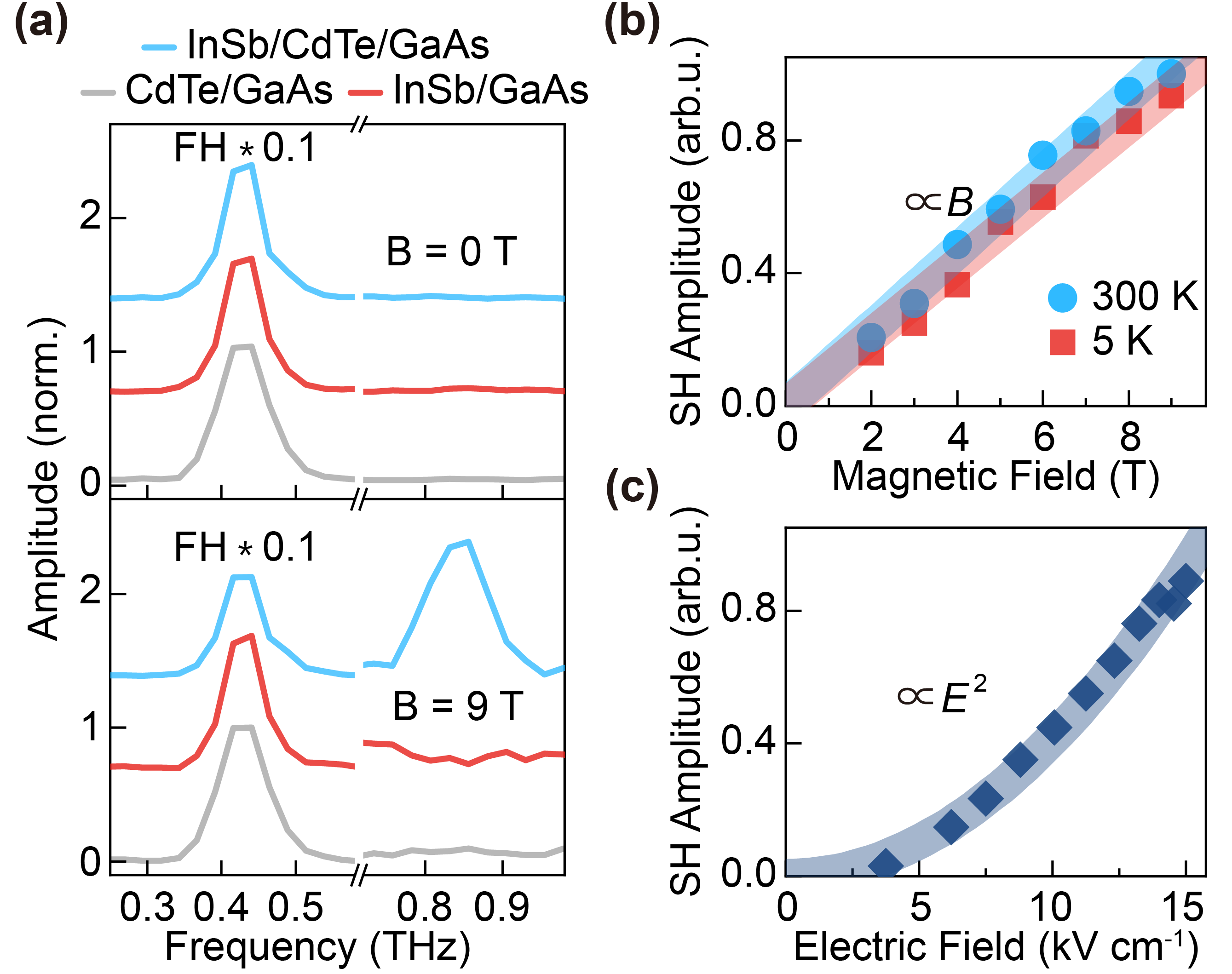}
  \caption{Magnetic-field-induced THz SHG in InSb/CdTe/GaAs.
    (a) Under identical excitation at 0.42 THz, no SHG is observed at zero magnetic field, whereas a pronounced SH signal at 0.84 THz emerges exclusively in InSb/CdTe/GaAs under a 9-T in-plane magnetic field. Curves are offset for clarity. (b,c) SHG amplitude scales linearly with the magnetic field strength ($B$) and quadratically with the incident electric field ($E^2$). }
  \label{fig:THz_SHG_magfield}
\end{figure}

To investigate the THz SHG mechanisms in the InSb/CdTe/GaAs sample, we conducted control experiments. As shown in Fig. \ref{fig:THz_SHG_magfield}(a) upper panel, none of the three samples exhibit a second-harmonic (SH) signal without an applied magnetic field. However, when an in-plane magnetic field is applied, the InSb/CdTe/GaAs sample demonstrates a pronounced SH signal [Fig. \ref{fig:THz_SHG_magfield}(a) lower panel]. In contrast, neither the InSb/GaAs nor the CdTe/GaAs samples show any SHG, even at 9 T. These results indicate that both the CdTe buffer layer and the application of an in-plane magnetic field are crucial for achieving THz SHG. The temporal waveform of the SH electric field is shown in Fig. \ref{fig:boat1}(b) (right axis). Notably, under moderate magnetic fields (e.g., 1~T), the THz SHG efficiency is orders of magnitude higher than that observed in other systems (SM Note 6).\cite{2023_Efficient_NC,2020_Subterahertz_S}  

Next, we explored the dependence of the THz SH signal on magnetic and electric fields in the InSb/CdTe/GaAs sample. By varying in-plane magnetic fields from -9 to 9 T, the FH signal exhibits negligible changes (Fig. S8). In contrast, the SH signal increases linearly with the field strength [Fig. \ref{fig:THz_SHG_magfield}(b)]. Notably, the SH signal persists up to room temperature, with slightly higher efficiency than at low temperatures. This room-temperature applicability suggests potential applications for THz devices under magnetic fields. In addition, when the magnetic field was reversed, the SH signal exhibited a clear $\pi$-phase shift (Fig. S9), while the transmitted FH signal retained its original phase (Fig. S8). Figure \ref{fig:THz_SHG_magfield}(c) demonstrates that the SH signal exhibits a quadratic dependence on the FH electric field and shows no saturation within the measurement range, unlike typical THz THG experiments.\cite{2018_Extremely_N,2020_Nonperturbative_NC,2020_Efficient_PRL}
This finding highlights the potential of local field enhancement in plasmonic nanostructure devices to achieve high THz SHG conversion efficiencies,\cite{2009_Terahertz_NP}
as well as to explore previously uninvestigated nonperturbative THz SHG effects.\cite{2021_GratingGraphene_AN} By contrast, the TH signal in our sample saturates at a similar incident field strength, which will be shown later.

\section{Symmetry analysis}
Given the established connection between nonlinear optical and transport effects and a material's point group symmetry, we performed a symmetry analysis to verify the compatibility of the observed nonlinear THz signal with the symmetry of the material and to clarify its microscopic origin.

In contrast to THz SHG, optical-frequency SHG is observed in all three samples (Fig. S6) at zero applied magnetic field, as anticipated due to the noncentrosymmetric nature of the $4\bar{3}m$ point group. Optical SHG occurs via (virtual) interband transitions, where a strong optical-frequency electric field induces a nonlinear polarization $P_i(2\omega)$ in a noncentrosymmetric material, characterized by $P_i(2\omega) = \chi^{(2)}_{ijk} E_j(\omega) E_k(\omega)$, with $\chi^{(2)}_{ijk}$ being a third-rank polar tensor. In comparison, THz SHG originates from the coherent intraband acceleration of electrons, which generates a nonlinear current. Our experimental observations indicate that this nonlinear current density can be described by $J_i(2\omega) = \sigma^{(3)}_{ijkl} E_j(\omega) E_k(\omega) B_l$, with $\sigma^{(3)}_{ijkl}$ being a fourth-rank axial tensor associated with the symmetry of the InSb/CdTe interface $3m$ or bulk InSb $4\bar{3}m$. 

\begin{figure}
    \centering
    \includegraphics[max width=\linewidth]{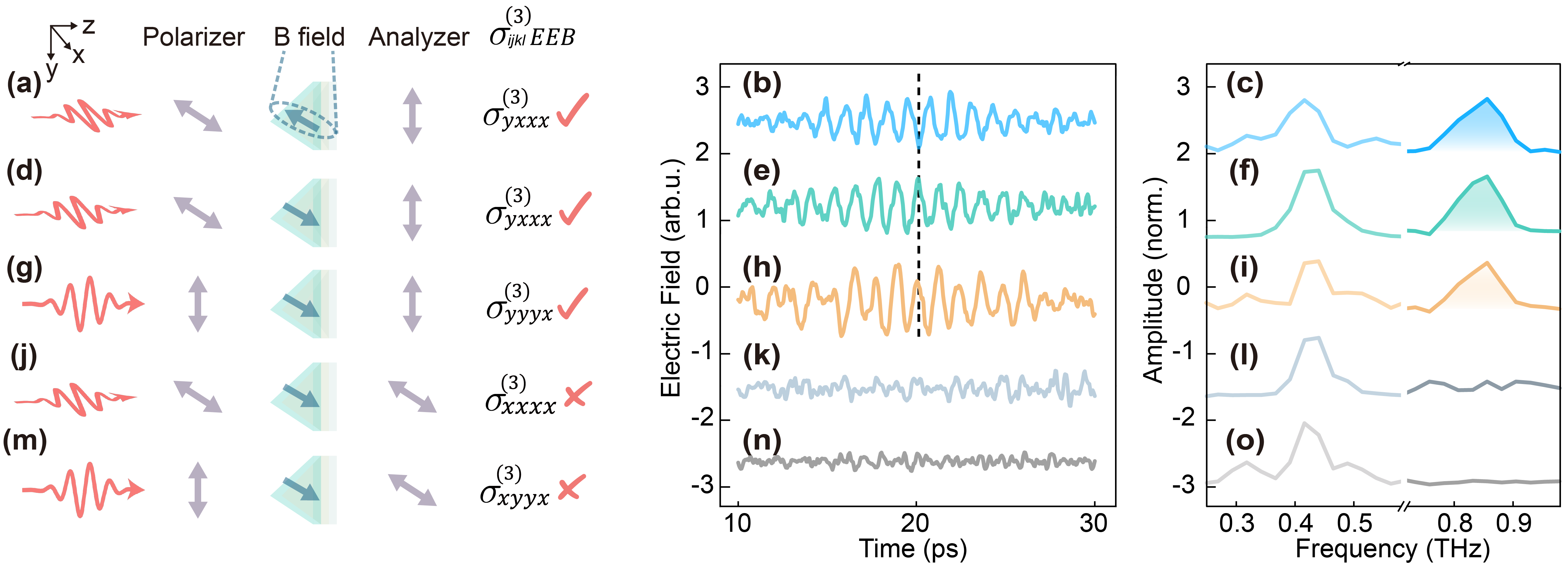}
    \caption{Symmetry analysis of magnetic-field-induced THz SHG. Measurement configuration and the corresponding time-domain trace and frequency domain spectrum for different tensor components.}
    \label{fig:THz_SHG_symmetry}
\end{figure}

To examine the symmetry constraints, we performed polarization-resolved THz SHG measurements, with the input and output polarizations controlled using motorized polarizers (SM Note 2). The sample was carefully mounted with the [-211] direction aligned with the magnetic field of the superconducting magnet (designated as the $x$-axis in Fig. \ref{fig:THz_SHG_symmetry}). The orientation of the sample was verified by the optical SHG experiment (SM Note 4). The THz wave propagates along the [111] direction (the $z$-axis). Figure \ref{fig:THz_SHG_symmetry} presents the experimental configuration in its first column, with the corresponding time- and frequency-domain signals illustrated in the second and third columns, respectively. 

In Fig. \ref{fig:THz_SHG_symmetry}(a-c), the SH signal corresponds to $J_y(2\omega) = \sigma^{(3)}_{yxxx} E_x(\omega) E_x(\omega) B_x$, which is symmetry-allowed for both the interface $3m$ and bulk $\bar{4}3m$ symmetry according to Neumann's principle (SM Note 8). When reversing the magnetic field from $-9$ to $+9$~T [Fig. \ref{fig:THz_SHG_symmetry}(d,f)], the THz SHG signal experiences a $\pi$-phase shift, as expected. By keeping the magnetic field and analyzer unchanged while varying the direction of the input electric field from the $x$-axis to the $y$-axis [in Fig. \ref{fig:THz_SHG_symmetry}(g-i)], the SHG signal persists. This configuration corresponds to $\sigma^{(3)}_{yyyx}$, which is also allowed by both $3m$ and $\bar{4}3m$. However, in Fig. \ref{fig:THz_SHG_symmetry}(j-o), no SHG is observed when the analyzer is aligned parallel to the magnetic field and the polarizer is aligned either parallel or perpendicular to the magnetic field, because the corresponding tensor elements $\sigma^{(3)}_{xxxx}$ and $\sigma^{(3)}_{xyyx}$ vanish under both $3m$ and $\bar{4}3m$. Our experimental results align well with both the interface and bulk symmetries, highlighting the potential for manipulating THz nonlinearity through crystalline symmetry engineering.

\section{Microscopic origins of THz SHG}
Analogous to our work in the THz regime, the nonlinear electric current response in the DC limit under the influence of an in-plane magnetic field has been extensively investigated both theoretically\cite{2021_Theory_PRBa,2023_Nonlinear_PRBb,2021_Anomalous_PRR,2023_Intrinsic_PRL,2023_Nonlinear_PRBc}
and experimentally.\cite{2023_RoomTemperature_AM,2017_Bulk_NP,2018_Observation_PRL,2025_Nonlinear_NMa,2019_Anomalous_PRL,2022_Gatetuneable_NM,2019_Nonlinear_NC,2020_Observation_PRL,2019_Strong_PRB} Given the parallels between nonlinear transport in the DC regime and our investigation of THz SHG, we proceed to examine the microscopic origins of THz SHG through theoretical calculations that extend previous studies into finite frequencies.\cite{2024_Unification_PRL,2024_Equivalence_PRB}

Our theoretical framework consists of two complementary components: (1) a model Hamiltonian describing a two-dimensional electron gas with Rashba SOC, representing the InSb/CdTe interface; and (2) first-principles calculations of bulk InSb, where the underlying physics can be viewed as that of a three-dimensional electron gas with Dresselhaus SOC. For each case, we employed semiclassical equations of motion for a Bloch electron wave packet, accounting for field modifications to the Berry curvature and band energy. By solving the Boltzmann equation and expanding the solution to the second order in the electric field and to the first order in the magnetic field, we obtained a nonlinear current response that exhibits a $J(2\omega) \propto E(\omega)^2 B$ dependence (SM Note 9). 

Contributions of various physical mechanisms were identified based on their power-law dependence on the relaxation time $\tau$. The $\tau^0$ term, which is independent of scattering, is referred to as the intrinsic term and originates from the material's band geometry. The $\tau^1$ term arises from the displacement current associated with polarization, which is negligible in our samples (SM Note 9.5). The $\tau^2$ term includes contributions from Zeeman coupling and Berry curvature effects, where the Zeeman part represents a magnetic-field-induced correction to the Drude conductivity.

\begin{figure}
    \centering
    \includegraphics[max width=\linewidth]{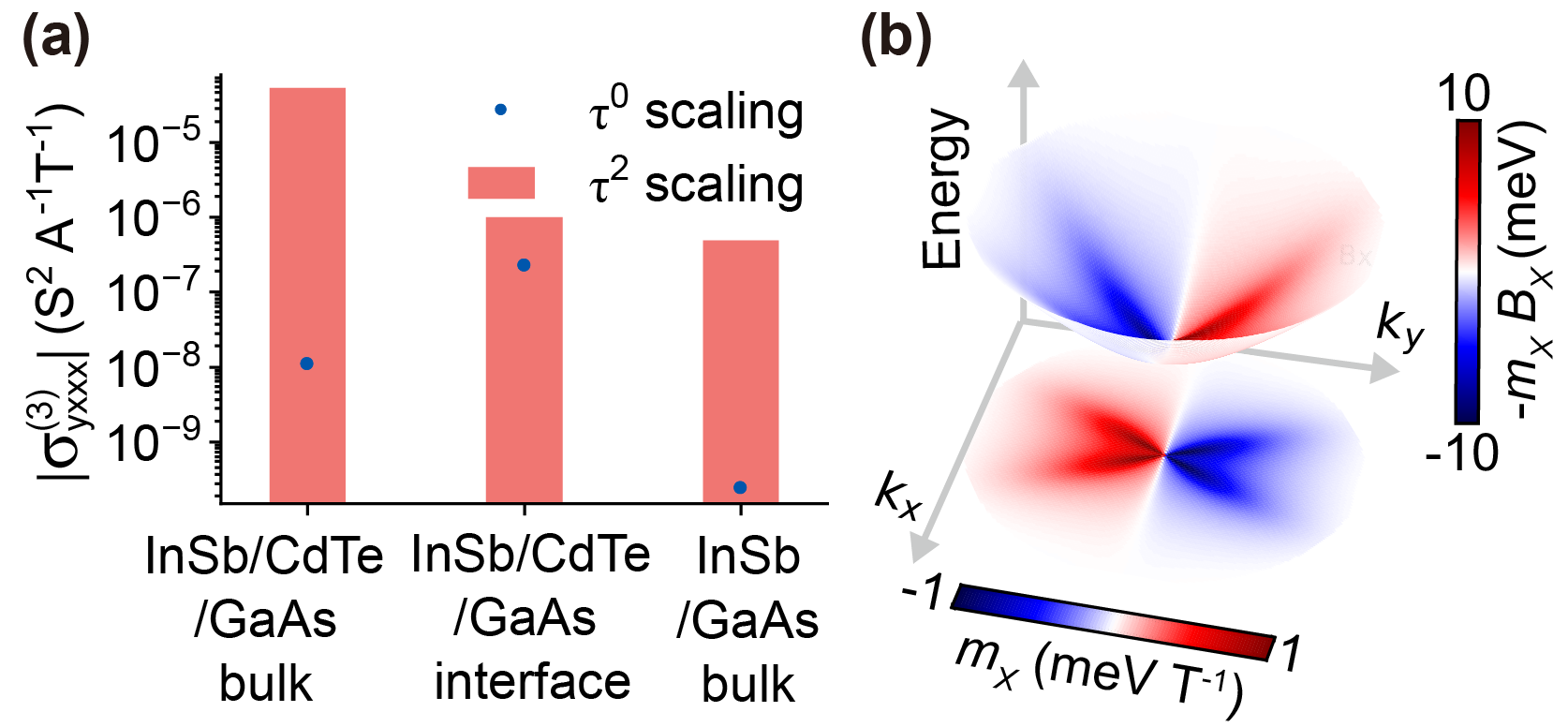}
    \caption{Microscopic SHG mechanism.
    (a) Magnitude of the effective nonlinear conductivity $|\sigma_{yxxx}^{(3)}|$ for various $\tau$-scaling contributions in different cases. (b) Orbital magnetic moment in momentum space for the upper Dresselhaus band of bulk InSb at $B=0$ (lower distribution) and the corresponding magnetic-field-modified band dispersion under $B_x=9$~T (upper surface).}
    \label{fig:THz_SHG_mechanism}
\end{figure}

Figure \ref{fig:THz_SHG_mechanism}(a) summarizes the calculated $|\sigma_{yxxx}^{(3)}|$ for the $\tau^0$- and $\tau^2$-scaling contributions in different cases.  A comparable trend is observed for $|\sigma_{yyyx}^{(3)}|$, as shown in Fig. S15. See SM Note 9.9 for details. In InSb/CdTe/GaAs, the dominant contribution arises from the $\tau^2$-term due to  the orbital Zeeman contribution in the bulk  sample, while the intrinsic $\tau^0$ term is several orders of magnitude smaller. 
In the following, we discuss the physical picture of magnetic-field-induced SHG due to the orbital Zeeman term.

In the absence of an external magnetic field, the original nonlinear Drude conductivity is proportional to $\sigma_{abc}=\tau^2\int [d\mathbf{k}]f_0' v_a \partial_{bc} \epsilon_\mathbf{k}$,  where $f_0' (\epsilon_\mathbf{k})$ is the derivative of the Fermi-Dirac distribution function with respect to energy, $v_a=\partial_{k_a} \epsilon_\mathbf{k}$ is the group velocity, and $\partial_{bc}=\partial^2_{k_bk_c}$ (SM Note 9.7). Since time-reversal symmetry requires the energy bands to satisfy $\epsilon_\mathbf{k}=\epsilon_\mathbf{-k}$, both $f_0' ( \epsilon_\mathbf{k})$ and $\partial_{bc} \epsilon_\mathbf{k}$ are even functions with respect to $\mathbf{k}$, whereas the group velocity $v_a(\mathbf{k})$ is an odd function of $\mathbf{k}$. Consequently, the integrand as a whole is an odd function with respect to $\mathbf{k}$, and its integral over the entire Brillouin zone must vanish. This demonstrates that in non-magnetic systems preserving time-reversal symmetry, the nonlinear Drude response is strictly forbidden.

The SOC due to inversion asymmetry produces a finite and asymmetric orbital magnetic moment, $\mathbf{m}(\mathbf{k}) \neq \mathbf{m}(-\mathbf{k})$. Time-reversal symmetry requires $\mathbf{m}(\mathbf{k}) = -\mathbf{m}(-\mathbf{k})\neq0$. Figure \ref{fig:THz_SHG_mechanism}(b) (lower 2D distribution) shows the calculated $m_x$ in the momentum space for the upper Dresselhaus band of bulk InSb at $\mathbf{B} = 0$.

Applying a magnetic field breaks time-reversal symmetry and introduces an additional Zeeman term, $-\mathbf{m} \cdot \mathbf{B}$ [Fig. \ref{fig:THz_SHG_mechanism}(b) upper 3D plot], to the band dispersion $\tilde\epsilon_\mathbf{k}=\epsilon_\mathbf{k}-\mathbf{m\cdot B}$. This modification leads to $\tilde\epsilon_\mathbf{k}\neq\tilde\epsilon_\mathbf{-k}$. As a consequence, both $f_0(\tilde\epsilon_\mathbf{k})$ and $\partial_{bc} \tilde\epsilon_\mathbf{k}$ are no longer even functions of $\mathbf{k}$. Additionally, the group velocity $v_a=\partial_{k_a} \tilde\epsilon_\mathbf{k}$ is no longer an odd function of $\mathbf{k}$. These changes prevent the cancellation of contributions with opposite momenta, resulting in a second-order current in the nonlinear Drude response. This provides a direct microscopic explanation for why both inversion-breaking (SOC) and time-reversal-symmetry breaking by the magnetic field are required for the observed THz SHG.

Compared with the bulk, the leading contribution to SHG from the InSb/CdTe interface also stems from the $\tau^2$ term, but is several times smaller [Figs. 4(a),S12]. At this two-dimensional interface, the orbital Zeeman and Berry curvature terms vanish. Instead, the relevant extrinsic contribution arises from the spin Zeeman term, which, like the bulk effect, requires both SOC and a magnetic field.

Furthermore, our calculations agree with the experimental results that the InSb/CdTe/\allowbreak{}GaAs sample exhibits significantly stronger SHG than the InSb/GaAs sample [Figs. 4(a),S13,S14]. This enhancement arises from two factors. First, the CdTe buffer layer increases carrier mobility, prolonging the relaxation time. As a result, the $\tau^2$-scaling orbital contribution is much larger in the InSb/CdTe/GaAs sample. Second, the lower Fermi level of the InSb/CdTe/GaAs sample results in a larger orbital magnetic moment, as shown in Fig. \ref{fig:THz_SHG_mechanism}(b), which in turn leads to a stronger SHG response.

\section{High-efficiency THz THG}

\begin{figure}
    \centering
    \includegraphics[max width=\linewidth]{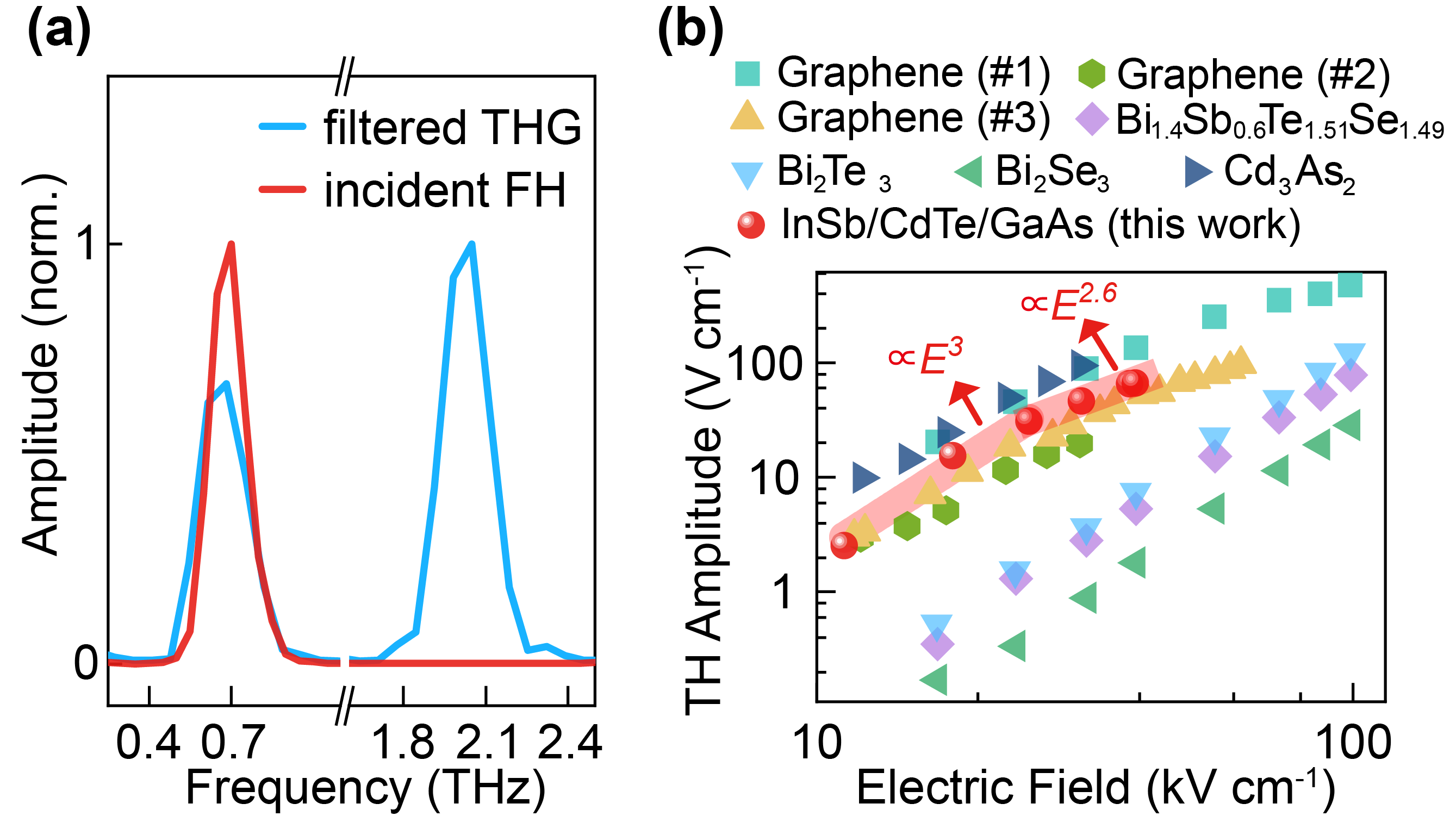}
    \caption{High-efficiency THz THG.
    (a) Spectra of the incident FH and generated TH in InSb/CdTe/GaAs without a magnetic field. (b) Comparison of THz THG efficiency across different material systems, showing high efficiency of our sample.}
    \label{fig:THz_THG}
\end{figure}

THz THG was measured without an applied magnetic field using incident THz pulses centered at 0.70~THz. Figures \ref{fig:boat1}(c) and \ref{fig:THz_THG}(a) display the time-domain waveform and frequency-domain spectrum of the generated TH signal, respectively. Figure \ref{fig:THz_THG}(b) presents the electric-field dependence of THG, along with prior THG results from various samples reported in the literature. The figure excludes (1) cases where the electric field strength is not explicitly specified,\cite{2024_THz_} and (2) research on surface field enhancement using metasurfaces.\cite{2009_Terahertz_NP,2025_Giant_a} The efficiency of our sample is comparable to benchmark THz THG materials such as Cd$_3$As$_2$ and graphene,\cite{2018_Extremely_N,2020_Nonperturbative_NC,2020_Efficient_PRL} and it is one order of magnitude higher than that of topological materials like Bi$_2$Se$_3$, Bi$_2$Te$_3$, and Bi$_{1.4}$Sb$_{0.6}$Te$_{1.51}$Se$_{1.49}$ (Ref.~\cite{2021_Terahertz_nQM}).

The THG signal is governed by $J_a(3\omega) = \sigma_{abcd}^{\mathrm{EEE}}E_bE_cE_d$, where $\sigma_{abcd}^{\mathrm{EEE}}$ is a fourth-rank polar tensor. Using finite-frequency nonlinear transport theory, we find that the THG primarily arises from a Drude-like intraband contribution in bulk InSb (SM Note 10), analogous to the magnetic-field-induced SHG. The small bandgap and low effective mass of InSb result in significant band nonparabolicity,\cite{1957_Band_JoPaCoS} and this, combined with high mobility, leads to a strong third-order response.\cite{1986_Farinfrared_PRBb} 

Figure \ref{fig:THz_THG}(b) illustrates that the TH signal in our sample saturates above 25~kV\,cm$^{-1}$; akin to behavior seen in materials with Dirac band dispersion.\cite{2018_Extremely_N,2020_Nonperturbative_NC,2020_Efficient_PRL} The saturation threshold in our thin film is ${\sim}0.01$~mJ\,cm$^{-2}$, at least an order of magnitude smaller than that of bulk single crystals.\cite{2024_Third_OLO} While the exact origin of this behavior remains unclear, potential factors include enhanced light-matter coupling due to Fabry--P\'{e}rot interference in the thin film and the relatively low excitation frequency employed here.

Importantly, while other THz THG-capable materials exhibit spatial inversion symmetry, the InSb/CdTe system does not, resulting in unique SHG and THG characteristics. The leading mechanisms for both SHG and THG are enhanced by relaxation time and require strong SOC. Consequently, high-mobility noncentrosymmetric semiconductors with strong SOC provide a promising pathway for achieving efficient THz harmonic generation beyond this specific sample. 

\section{Discussion}
In conclusion, we have demonstrated efficient SHG and THG in the InSb/CdTe heterostructure at THz frequencies. Our work represents, to the best of our knowledge, the first illustration of a THz-frequency magnetic field-induced nonlinear transport response in a semiconductor heterostructure. Unlike widely studied magnetic-field-induced SHG in the optical range,\cite{2005_MagneticFieldInduced_PRL,2004_MagneticField_PRL} which is dominated by interband transitions, our results shed light on the nonlinear response near the Fermi surface. This all-optical, contact-free approach provides sub-picosecond time resolution, making it ideal for time-resolved experiments that explore ultrafast THz nonlinear dynamics\cite{2009_Impact_PRB,2023_Ultrafast_PRL} and ultrafast manipulation of novel quantum states. Furthermore, our work can be extended to the exploration of nonlinear transport phenomena, such as the nonlinear Hall effect\cite{2021_Nonlinear_NRP} and nonlinear transport in high Landau level systems.\cite{2012_Nonequilibrium_RMP} The use of ultrashort and strong THz pulses enables the investigation of nonperturbative nonlinear responses that are inaccessible to low-frequency AC transport measurements due to heating. 

To facilitate practical application, we propose engineering techniques to eliminate the need for an external magnetic field, such as incorporating a ferromagnetic layer or leveraging the magnetic proximity effect.\cite{2022_GateTunable_NL,2025_Quantum_NC} Furthermore, integrating the heterostructure with THz metamaterial field-enhancement structures could significantly reduce the required magnetic field while amplifying the electric field in the sample by orders of magnitude.\cite{2009_Terahertz_NP} Our high-quality epitaxial thin-film platform offers scalability, stability, and electrical controllability, suggesting its potential for device integration.\cite{2021_NarrowBand_2IIEDMI}

\section*{Acknowledgments}

\begin{acknowledgments}
We thank Zhiyuan Sun for helpful discussions. The work was supported by the National Natural Science Foundation of China (Grants No.~12421004), the National Key R\&D Program of China (Grants No.~2021YFA1400100, No.~2025YFA1411200, and No.~2024YFA1611300), the National Natural Science Foundation of China (Grants No.~12361141826, No.~12074212 and No.~11974414), and the Beijing Natural Science Foundation (Grants No.~Z240006). This work was supported by the Synergetic Extreme Condition User Facility (SECUF, \url{https://cstr.cn/31123.02.SECUF}).
\end{acknowledgments}

\bibliography{refs}

\end{document}